\documentstyle[aps,amssymb]{revtex}
\title{Optical Fock-state synthesizer}
\author{G. M. D'Ariano, L. Maccone, M. G. A. Paris, and M. F. Sacchi}
\address{Theoretical Quantum Optics Group, 
INFM and Dipartimento di Fisica ``A. Volta'' \\
Universit\`a di Pavia \\ Via A. Bassi 6, I-27100 Pavia, Italy}
\begin{document}
\date{\today}\maketitle
\begin{abstract}We suggest a tunable optical device to synthesize Fock
states and their superpositions starting from a coherent source. The
scheme involves an avalanche triggering photodetector and a ring cavity
coupled to a traveling wave through a cross-Kerr medium. Low quantum
efficiency at the photodetector improves the synthesizing quality at the
expense of reducing the synthesizing rate.\end{abstract}
\section{Introduction}
In the last decade, much attention has been devoted to the quantum
engineering of nonclassical light. Different types of nonclassical
states can be now prepared, and their quantum properties can be
entirely characterized by accessible measurement schemes. Among the
various quantum states of the optical field, states with a definite
number of photons deserve a special attention. Indeed, besides the
interest for fundamental tests of Quantum Mechanics \cite{pha}, the
Fock (number) states are also relevant for many applications, as for
example, to achieve the optimal capacity coding in quantum
communication channels \cite{dru}, or the Heisenberg sensitivity limit
in high-precision quantum interferometry \cite{hol}. \par Different
methods for the generation of Fock states have been proposed, both for
traveling-wave and cavity fields. For traveling-wave fields, these
methods are principally based on tailored nonlinear interactions
\cite{non}, conditional measurements \cite{con}, or state engineering 
\cite{eng}. The experimental realizability or effectiveness of these 
proposals is often challenging. On the other hand, Fock states have
been generated into cavities \cite{cav}, i.e. using the micromaser
trapped states. \par In this paper we address the problem of Fock
state generation in a traveling mode.  We suggest an optical device
based on an avalanche triggering photodetector and a ring cavity
coupled to an external traveling wave through a cross-Kerr
medium. Remarkably, the input states of our proposal are just
customary coherent states. Our scheme differs from conventional setups
involving conditional photon counters \cite{con}, since we use simple
on--off detection of an intense coherent field, allowing very low
quantum efficiency at the photodetector. Moreover, as we will see, the
scheme can also be used to engineer superpositions of few Fock states,
which are a crucial resource for optical quantum computers
\cite{com}, and quantum tomography of optical Hamiltonians \cite{ham}.
\par The main advantages of the present scheme are its tunability in
preparing any chosen number state and selected superpositions, along
with its robustness against the imperfections of the triggering
photodetector.  Indeed, low quantum efficiency at the photodetector
has no detrimental effect on the filtering process, but only reduces
the state-synthesizing rate. \par The paper is structured as
follows. In the next Section the scheme is introduced, and its use in
preparing Fock states is described. In Section \ref{s:sup} we show how
the same scheme can be used to prepare superpositions of few Fock
states, whereas in Section \ref{s:eta} the effects of non-unit quantum
efficiency at the photodetector are taken into account. Section
\ref{s:outro} closes the paper by summarizing results, and discussing
the feasibility of the scheme.
\section{Synthesis of Fock states}
The schematic setup of the synthesizer is depicted in Fig. \ref{f:experiment}.
The ring cavity is build by two mirrors and two high reflectivity beam 
splitters. Here, for simplicity, we suppose that the beam splitters have 
the same transmissivity $\tau$. The cavity is fed by a coherent state in 
the mode $a_1$, whereas the mode $a_2$ is left unexcited. Through the 
cross-Kerr interaction, the cavity mode $d$ is coupled to an external 
traveling mode $c_1$, according to the unitary evolution \cite{ker}
\begin{eqnarray}
\hat U_{K}=\exp(-i\chi t \,d^\dag d \,c_1^\dag c_1)
\;,\label{ukerr}
\end{eqnarray}
$\chi$ being the nonlinear susceptibility of the medium and $t$ the
interaction time. The signal mode $c_1$ is prepared in a coherent
state, and, additionally, a tunable phase shift $\psi$ is introduced
in the cavity mode. \par The scheme is suitable for applications with
both continuous waves and pulses. In a situation involving pulses,
however, in order have states with a definite number of photons we
need a long enough quantization time. Indeed, the coherence time of
the signal mode should be quite long compared to the probe one. We
assume that the coherence time of the input signal is of the order of
the photon flight time in the cavity, thus assuring that the cavity
mode effectively couples with the signal through the cross-Kerr
medium. Moreover, the coherence time of the cavity mode should be
shorter than the signal one, thus allowing the Kerr phase to cumulate
in the loop. In summary, the coherence time of the probe mode should
be shorter than the signal one, which, in turn, is determined by the
photon flight time in the cavity.  \par At the output of the cavity
the field is monitored by an avalanche photodetector. For the purpose
of our scheme, we only need to verify the presence or absence of the
field, at the output port of the cavity through the triggering
photodiode D. Let us initially assume unit quantum efficiency at
photodetection. The measurement is described by the two-value
probability operator measure (POM) $\hat \Pi _n$
\begin{eqnarray}
\hat\Pi_0\doteq|0\rangle\langle 0|\;, \qquad \hat\Pi_1\doteq \hat I_{b_2}-|0
\rangle\langle 0|\label{pom}\;,
\end{eqnarray}
where $|0\rangle $ is the vacuum and $\hat I_{b_2}$ is the identity for mode
$b_2$.  As we will show in the following, due to the very steep
dependence of the cavity transmissivity on the total phase 
shift---including both cross Kerr interaction and phase $\psi$---the
detection of the field at photodetector D guarantees that the free
mode $c_2$ at the output of the Kerr medium is reduced into a Fock
state or a superposition of few Fock states.  \par 
The mode transformations of the ring cavity are \cite{lou}
\begin{eqnarray}
\left\{\begin{array}{l}b_1=\kappa(\varphi)
a_1+e^{i\varphi}\sigma(\varphi) a_2\\ b_2=\sigma(\varphi) 
a_1+\kappa(\varphi) a_2\end{array}\right.
\;,\label{ring}
\end{eqnarray}
where the phase-dependent cavity transmissivity $\sigma$ and
reflectivity 
$\kappa$ are given by
\begin{eqnarray}
\kappa(\varphi)&\doteq&\frac{\root\of{1-\tau}(e^{i\varphi}-1)}
{1-e^{i\varphi}(1-\tau)}\nonumber \\
\sigma(\varphi)&\doteq&\frac\tau{1-
e^{i\varphi}(1-\tau )}\label{defsigma}
\end{eqnarray}
with $\left|\kappa(\varphi)\right|^2+\left|\sigma(\varphi)\right|^2
=1$. The transformations (\ref{ring}) and (\ref{defsigma}) are
rigorously obtained by quantizing the e.m. field modes which solve 
the Helmholtz equation of the etalon, as in Ref. \cite{lou},
and taking the input/output modes of the asymptotic free plane
waves. However, a {\em naive} solution of the etalon as a loop of beam
splitters gives the same result, with the internal modes having a
reduced commutator (this point is well explained in Ref. \cite{bar}).
For $c_1$ in the Fock state $|n\rangle$, the total phase shift is given
by 
\begin{eqnarray}
\varphi=\psi - \chi n t\equiv \varphi _n
\;.\end{eqnarray}
To simplify the notation, 
we write $\sigma _n\doteq \sigma(\varphi _n)$ and analogously for
$\kappa$. Let us now consider the input state
\begin{eqnarray}
\hat\varrho_{in}=|\alpha\rangle\langle 
\alpha|\otimes|0\rangle\langle 0|\otimes\hat\nu_{in}
\;,\label{rhoin}
\end{eqnarray}
namely a generic state $\hat\nu_{in}$ for mode $c_1$, a
coherent state $|\alpha\rangle$ for mode $a_1$, and vacuum for
$a_2$. In the Schr\"odinger picture the output state can be written
in the form
\begin{eqnarray}
\hat\varrho_{out} = \label{rhoout} 
\sum_{n,m=0}^{\infty} \nu_{nm} \: |\kappa_n\alpha\rangle
\langle \kappa_m\alpha|\otimes|\sigma_n\alpha\rangle\langle
\sigma_m\alpha|\otimes |n\rangle\langle m|\;.
\end{eqnarray} 
The process of filtering the desired Fock state from the input state 
$\hat\nu_{in}$ is triggered by the photodetector D as follows. 
The probabilities corresponding to the outcomes $1$ (detector D on)
and $0$ (detector D off) are given by
\begin{eqnarray}
P_1 = \hbox{Tr} \left[\hat\varrho_{out}\: \hat\Pi_1 
\right] = \sum_{n=0}^\infty \nu_{nn} \left(1-
e^{-|\alpha|^2\:|\sigma_n|^2}
\right)
\label{probs}\;,
\end{eqnarray}
and $P_0=1-P_1$. 
By means of Eq. (\ref{defsigma}) we have
\begin{eqnarray}
|\sigma_n|^2 = \left(1+ 4 \frac{1-\tau}{\tau^2}\sin^2 
\frac{\psi-\chi nt}{2} \right)^{-1}
\label{sigma}\;,
\end{eqnarray}
which is a periodic function sharply peaked at
\begin{eqnarray} 
n=\frac{\psi +2\pi j}{\chi t}\doteq n^*+\frac{2\pi}{\chi t}j \;,\qquad
j\in{\mathbb Z}    
\;,\label{condn}  
\end{eqnarray}  
with unit maximum height and width of the same order of the 
beam splitter transmissivity 
$\tau$ (typically $\tau\sim 10^{-4}\div 10^{-6}$).  The value $n^*$ 
in Eq. (\ref{condn}) can be adjusted to an arbitrary integer by 
tuning the phase-shift $\psi$
as a multiple of $\chi t$, whereas multiple resonances are avoided 
by using small nonlinearity $\chi t \ll 1 $, so that the
values of $n$ satisfying Eq.  (\ref{condn}) for $j\not = 0$ correspond
to vanishing matrix elements $\nu_{ni}\simeq 0\ \forall i$. 
In this case for high-quality cavities $\tau\ll \chi t $ in
Eq. (\ref{probs}) we have $|\sigma_n|^2 \simeq\delta_{nn^*}$, and
the detection probability $P_1$ rewrites
$P_1 \simeq \nu_{n^* n^*} \left(1- e^{-|\alpha|^2
}\right)$. Notice that increasing the amplitude of $\alpha$ will 
enhance the detection probability $P_1$. Moreover, 
one can optimize also the input state $\hat\nu_{in}$ to achieve the
highest $\nu_{n^*n^*}$. For example, in the case of a coherent input
$\hat\nu_{in}=|\beta\rangle\langle\beta|$, one could select
$|\beta|\simeq \sqrt{n^*}$. 
\par We now evaluate the conditional state
$\hat\nu_{out}$ at the output of the Kerr medium for detector D
on. One has 
\begin{eqnarray}
\hat\nu_{out} =\frac{1}{P_1}\: \hbox{Tr}_{a_1\; a_2} 
\left[ \hat\varrho_{out} \:\hat\Pi_1 \right]= 
\frac{e^{-|\alpha|^2}}{P_1} 
\sum_{n,m=0}^\infty \nu_{nm}\: e^{|\alpha|^2 \kappa_n\kappa^*_m}\: 
\bigg(e^{|\alpha|^2 \sigma_n\sigma^*_m} -1\bigg)\:|n\rangle\langle  m|
\;, \label{strainer2} 
\end{eqnarray}
where the partial trace is performed over the ring cavity modes.  The
argument $\theta $ of $\sigma(\varphi )=|\sigma(\varphi )| \exp
[i\theta(\varphi)]$ is given by $\theta (\varphi) = \arctan\left[
\frac{(1-\tau)\sin \varphi }{1-(1-\tau)\cos \varphi}\right]$. For
$\tau\ll 1$, as already noticed, $|\sigma_n|$ is nonzero only for
$n=n^*$, and correspondingly we have $\theta (\varphi _n)=0$.
Therefore, for all practical purposes we can write $\sigma_n
\sigma^*_m \simeq |\sigma_n| \:|\sigma^*_m| \simeq \delta_{nn^*}
\delta_{mn^*}$, and the output state (\ref{strainer2}) becomes
\begin{eqnarray}
\hat\nu_{out} \simeq |n^* \rangle\langle  n^*|\;,\qquad \tau\ll \chi t
\label{s3}\;,
\end{eqnarray}
{\it i.e .} the Fock component $|n^*\rangle$ has been filtered from the
initial state $\hat\nu_{in}$. \par
In Fig. \ref{f:diag} we report the number distribution of the
conditional output state $\hat \nu_{out}$, with $\psi$ tuned to obtain
$|n^*=4\rangle$ for different values of the beam splitter 
transmissivity.  
\section{Synthesis of superpositions}\label{s:sup}
We now show how the same setup may be used to produce
superpositions of Fock states. By choosing higher nonlinearities, the
quantity $2\pi/ (\chi t)$ decreases and $|\sigma_n|^2$ can be
significantly different from zero for more than one value of $n$ 
corresponding to sizeable components of the input state $\hat\nu_{in}$. If
there are only two of these ``resonant'' values 
$n_1=n^*$ and $n_2=n^*+2\pi/\chi t$, we have $|\sigma_n|\simeq
\delta_{nn_1}+\delta_{nn_2}$ and $P_1$ now reads
\begin{eqnarray}
P_1 \simeq(\nu_{n_1 n_1} +
\nu_{n_2 n_2})\left(1-e^{-|\alpha|^2}\right)\;,\qquad\tau\ll\chi t 
\label{lastprobs}\;.
\end{eqnarray}
Accordingly, the conditional state after a successful photodetection
becomes
\begin{eqnarray}
&&\hat\nu_{out}\simeq\frac 1{\nu_{n_1 n_1} +
\nu_{n_2 n_2}}
\Big[
\nu_{n_1 n_1}|n_1 \rangle\langle  n_1| +
\nu_{n_2 n_2}|n_2 \rangle\langle  n_2|  \nonumber \\
&+& \nu_{n_1 n_2}|n_1 \rangle\langle  n_2| +
\nu_{n_2 n_1}|n_2 \rangle\langle  n_1|
\Big]\;\qquad \tau\ll \chi t \;,
\label{s4}
\end{eqnarray}
which is a pure state if and only if $\nu_{n_1 n_1}\nu_{n_2
n_2}=\nu_{n_1 n_2}\nu_{n_2 n_1}$, namely for $\hat\nu_{in}$ in a pure 
state. In Fig. \ref{f:off} we report the density matrix of the
conditional output state $\hat \nu_{out}$, with $\psi$ tuned to obtain
a superposition of the Fock states $|n_1 \equiv10\rangle$ and
$|n_2\equiv 20 \rangle$ for different values of the beam splitter
transmissivity.  It is worth noting that the coefficients of the
superposition in Eq.  (\ref{s4}) are selected by the input state
$\hat\nu_{in}$. Therefore, in order to have a superposition with equal
weights starting from a coherent state $|\beta\rangle $, it is
sufficient to choose its amplitude in such a way that 
$|\beta|^2 = \left({n_1 !}/{n_2 !}\right)^{1/(n_1-n_2)} $. 
Notice, however, that with this choice of $\beta$ we find a small
contribution due to the term with $j=2$ in Eq. (\ref{condn}).
\section{Effects of imperfect photodetection}\label{s:eta}
Let us now take into account the quantum efficiency $\eta$ at the 
photodetector. In this case the POM $\hat\Pi_n$ is
replaced with 
$\hat\Pi^{(\eta)}_n$, where
\begin{eqnarray}
\hat\Pi_0^{(\eta)}\doteq\sum_{k=0}^\infty(1-\eta)^k|k\rangle\langle
k| \;,\qquad \hat\Pi_1^{(\eta)}\doteq I_{b_2}-\hat\Pi_0^{(\eta)} 
\label{pometa}\:.
\end{eqnarray} 
The probability $P_1^{(\eta)}$ of having outcome $1$ at 
the photodetector and the conditional output state now are the
following 
\begin{eqnarray} 
&&P_1^{(\eta)}= \mbox {Tr}[\hat\Pi_1^{(\eta)}\hat\varrho_{out}] =
\sum_{n=0}^\infty \nu_{nn} \left(1- e^{-\eta|\alpha|^2\:|\sigma_n|^2}\right)
\;,\label{strainereta0} \\
&&\hat\nu_{out}^{(\eta)}=
\frac{e^{-|\alpha|^2}}{P_1^{(\eta)}} 
\sum_{n,m=0}^\infty \nu_{nm}\: e^{|\alpha|^2 [\kappa_n\kappa^*_m
+\sigma_n\sigma^*_m] } 
\bigg(1-e^{-\eta|\alpha|^2 \sigma_n\sigma^*_m}\bigg)\:|n\rangle\langle  m|
\;.\label{strainereta1}
\end{eqnarray}
Remarkably, low values for the quantum efficiency $\eta$ can reinforce
the process of filtering, though at the cost of lowering the
probability $P_1^{(\eta)}$ of photodetection. In fact, $\eta$ scales
the term $\sigma_p\sigma^*_q$ in the exponential in
Eq. (\ref{strainereta1}), thus lowering the off-resonance
contributions. In Fig. \ref{f:offeta}a we actually purify the superposition 
shown in Fig. \ref{f:off}a by lowering the quantum efficiency to the value 
$\eta=20\%$. The probability of obtaining the state is correspondingly 
lowered from the value $P_1=0.205$ to $P_1^{(\eta)}=0.116$. An analogous 
argument holds for the dependence of the output state $\hat\rho_{out}$ on 
the input coherent amplitude $\alpha$. In Fig. \ref{f:offeta}b we report the
conditional state obtained by choosing $\alpha=3.58$, which corresponds to the 
same detection probability $P_1^{(\alpha)}=0.116$.
Obviously the above discussion on the effect of non-unit quantum
efficiency and on the intensity of the input coherent state
$|\alpha\rangle$ holds also for the generation of single Fock states.
\section{Conclusions and outlook}\label{s:outro}
The present proposal is characterized by two relevant features: its
tunability in the preparation of any chosen number state and of
selected superpositions, and its robustness against the imperfections
of the photodetection process. Therefore, it is a matter of interest
to discuss its feasibility in practical applications. This mostly
depends on two parameters: the quality factor of the ring cavity
(governed by the beam splitters transmissivity $\tau$) and the value
of the nonlinear coupling $\chi t$. For a good cavity (i.e. a cavity
with large quality factor) $\tau$ should be quite small, usual values
achievable in quantum optical labs being of the order of $\tau \simeq
10^{-4}$--$10^{-6}$. Remarkably, losses due to absorption processes
are of the order of $10^{-7}$, at least one order of magnitude smaller
than $\tau$. The scheme also requires an appreciable value for the
nonlinear cross-Kerr coupling between the signal and the cavity
modes. The coupling can be realized by optical nonlinear couplers, either 
codirectional or contradirectional, in the case of an all-optical fiber
implementation, or by direct matching of modes on a third-order crystal
in the case of a free field. 
Usually, third-order nonlinearities are small, and may be masked by
the concurrent self-modulation and absorption processes. Although the
effects of self phase modulation can be avoided using resonant
$\chi^{(3)}$ media \cite{ker}, currently available nonlinearities can
be used in the present scheme only for the preparation of number
states \cite{nota}.  However, recent breakthroughs in nonlinear Kerr physics
\cite{3turq,behr} open new perspectives for applications in quantum
optics. In fact, methods based on dark atomic resonance and
electromagnetically induced transparency have been suggested to
strongly enhance nonlinearity while suppressing absorption. These
results (especially for atomic gases) indicate that giant Kerr nonlinear
shifts of the order of 1 radiant per photon may be obtained by methods 
not too far from present technology \cite{behr,add}. \par 
As regards the power of the output signal (and the probe mode), we mention
that this should not be too large, since the $2\pi$-periodic resonance 
structure of the cavity transmissivity relies on the linear phase-shift
imposed to the cavity mode. If the power is too large the nonlinear 
contribution drastically alters this situation, and gives rise to bistability
or period-doubling instabilities.  The actual value of the allowed power
strongly depends on the specific nonlinear crystal employed in the
setup. We do not discuss here the technical details.\par In
conclusion, we have suggested a novel scheme to generate arbitrary
optical photon-number eigenstates in a traveling wave mode. The scheme
uses on--off photodetection of the field mode exiting a high-Q cavity,
which, in turn, is coupled to the traveling-wave by nonlinear Kerr
interaction.  The input fields for the setup are just customary
coherent states. After a successful photodetection, the traveling mode
is found in a photon-number eigenstate, or, for sufficiently high Kerr
nonlinearity, in a superposition of Fock states. The photodetector is
needed just to test the presence of an intense coherent output field
and, in fact, we have shown that non-unit quantum efficiency at
photodetection improves the quality of the state synthesis, however at
the expenses of the synthesizing rate.
 
\begin{figure}
\caption{Schematic diagram of the setup for the generation of Fock states 
and superposition of Fock states. BS$_1$ and BS$_2$ denote high
transmissivity beam splitters, and $\psi$ a tunable phase shift. The
cavity input modes $a_1$ and $a_2$ are in a coherent state and in the
vacuum respectively. The box denotes the cross Kerr medium that
couples the cavity mode $d$ with the traveling free mode $c_1$ which
is initially prepared in a coherent state. After a successful
detection at the photodiode D, which tests the presence of an output
excited coherent field in mode $b_2$, the output mode $c_2$ is found
in a Fock state or in a superposition of few Fock states.}
\label{f:experiment}
\end{figure}
\begin{figure}
\caption{The number distribution of the conditional output state $\hat 
\nu_{out}$ for different values of the BS transmissivity $\tau$, indicated 
on each plot. The parameters are chosen to select the number state $|n^* 
\equiv 4 \rangle$, with $\chi t = 0.01$ and $\psi=0.04$. The modes $a_1$ 
and $c_1$ are both excited in a coherent state with real amplitude $\alpha 
=20 $ and $\beta =2 $, respectively. The probabilities of obtaining the 
three states (i.e. the detection probability at D) are respectively $P_1
=0.99885,\ 0.4905,\ 0.1997.$} \label{f:diag}
\end{figure} 
\begin{figure}
\caption{The density matrix of the conditional output state $\hat 
\nu_{out}$ for different values of the BS transmissivity.
The parameters are chosen
to select the superposition $2^{-1/2}(|10\rangle + |20\rangle)$, that
is $\chi t = \pi/5$ and $\psi=0$. The modes $a_1$ and $c_1$ are both
excited in a coherent state with amplitude $\alpha =8.0 $ and $\beta =
(20!/10!)^{1/20}\simeq 3.902$, respectively. The detector has unit
quantum efficiency. Notice the small components due to the Fock state
$|30\rangle$. The probabilities of synthesizing these two states are
$P_1=0.205$ and $0.092$, respectively.}
\label{f:off}
\end{figure}
\begin{figure}
\caption{The density matrix of the conditional output state $\hat 
\nu_{out}$ as in Fig. \ref{f:off}a. Here we show the matrix elements 
for lowered quantum efficiency $\eta=20\%$ in (a), and for lowered 
input amplitude $\alpha=3.58$ in (b). Notice, in both cases, the 
improvement over the reconstructed state of Fig. \ref{f:off}a.}
\label{f:offeta}
\end{figure}

\begin{thebibliography}{99}
\bibitem{nota} As an order of magnitude for the parameters involved,
consider that an electric field $E\sim 10^2$ V/m is achievable for a
meter long $\sim 10$ photon pulse with a beam cross section of $\sim
10^{-10}$ m$^2$, corrisponding to a pulse power per unit area of 1
mW/cm$^2$ and a frequency $\omega\sim 10^{15}$ Hz (see \cite{behr}).
\end{thebibliography}
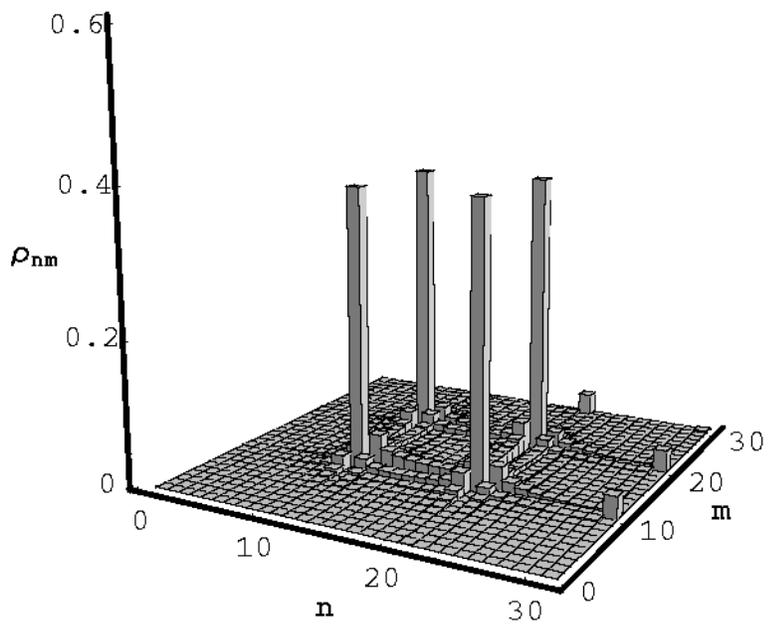
\end{document}